\documentstyle[aps,multicol, psfig]{revtex}
\renewcommand{\narrowtext}{\begin{multicols}{2} \global\columnwidth20.5pc}
\renewcommand{\widetext}{\end{multicols} \global\columnwidth42.5pc}

\def\nb #1{{\hbox{\bf #1}}}

\def\bOmega{\overline{\Omega}}
\def\oob{\Omega \bOmega}
\def\emn{\varepsilon_{\mu\nu}}
\def\nbm{\nb n}
\def\nm{n}

\begin{document}

\title{Scattering of Magnetic Solitons in two dimensions}
\author{S. Komineas}
\address{Physikalisches Institut, Universit\"at Bayreuth,
D-95440 Bayreuth, Germany \\
e-mail: stavros.komineas@uni-bayreuth.de}
\maketitle

\begin{abstract}
Solitons which have the form of a
vortex-antivortex pair have recently
been found in the Landau-Lifshitz equation which is the standard
model for the ferromagnet. We simulate numerically head-on
collisions of two vortex-antivortex pairs and observe
a right angle scattering pattern.
We offer a resolution of this
nontrivial dynamical behaviour by examining the Hamiltonian
structure of the model, specifically the linear momentum
of the two solitons.
We further investigate the dynamics of vortices in a modified
nonlinear $\sigma$-model which arises in the
description of antiferromagnets. We confirm numerically that a
robust feature of the dynamics
is the right angle scattering of two vortices which collide head-on.
A generalization of our theory is given for
this model which offers arguments towards
an understanding of the observed dynamical behaviour.
\end{abstract}


\narrowtext

\section{Introduction}

Localized solutions, often called solitons, play an increasingly
important role in nonlinear field theories 
in two dimensions. 
Topological structures exist in particular in magnetic systems and have been
studied extensively, both theoretically and experimentally 
\cite{slonc,baryak}.

An easy-plane ferromagnet is described by the
Landau-Lifshitz equation
\begin{equation}
\label{eq:lle}
 \dot{\nbm} = {\nbm} \times {\nb f},
\end{equation}                                                                  
$$
 {\nb f} = \Delta {\nbm} - \nm_3\, \hat{\nb e}, \qquad
 \nbm^2 = 1.
$$
The field $\nbm$ represents the local magnetization of the material.
We shall study here the case of a two-dimensional medium so we assume 
$\nbm \!=\! \nbm(x,y,t)$. The dot denotes a time derivative,
$\nm_3$ is the third component of $\nb n$, 
$\Delta$ is the Laplace operator and $\hat{\nb e} \!=\! (0,0,1)$ is the
unit vector in the third direction. 
The normalization condition $\nbm\!=\!1$ imposed in the initial
condition is preserved by the equations of motion.

Static solutions of model (\ref{eq:lle}) are the well-studied
vortices. Isolated vortices are spontaneously 
pinned objects that is
no vortex in free translational motion can be found 
in a 2D ferromagnet.
The same is true for any
isolated object with nontrivial topology in a 2D ferromagnet \cite{pt}, 
the most well-known example
being the magnetic bubbles in easy-axis ferromagnetic films \cite{slonc}.

Coherently traveling solutions of (\ref{eq:lle}) have been found in \cite{semi}.
They have the form of a vortex-antivortex pair and
their velocity may take any value 
between zero and unity, which is the velocity
of magnons in the system.

We now turn to a different class of systems, namely 
antiferromagnets. The dynamics of the staggered magnetization
in the antiferromagnetic continuum is given by the nonlinear $\sigma$-model
\cite{baryak,halperin,afm}:
\begin{equation}
\label{eq:sigma}
 \nbm \times [\,\ddot{\nbm} - {\nb f}\,] = 0,
\end{equation}
$$
 {\nb f} = \Delta \nbm - \nm_3\, \hat{\nb e}, \qquad
 \nbm^2 = 1,
$$                                                                              
where the double dot denotes a second time derivative.
The above model has the same static vortex solutions as (\ref{eq:lle}).
On the other hand, vortices in model (\ref{eq:sigma}) can be found
in free translational motion. This is due to the fact that the model
is invariant under Lorentz transformations.

Our main purpose is to study collisions of solitons in models
(\ref{eq:lle}) and (\ref{eq:sigma}).
In the ferromagnet no collision between two vortices can take place.
Two vortices with the same topological charge
will rotate around each other while a vortex and an
antivortex undergo Kelvin motion perpendicular to the line
connecting them.
However, collisions can occur between two vortex-antivortex pairs.
We elaborate on the
arguments of \cite{semi,cooper} and argue that head-on collisions between
vortex-antivortex pair solitons give a right angle scattering pattern.

We also study collisions between vortices in
antiferromagnets in Eq.~(\ref{eq:sigma}).
They scatter at right angles as found in numerical simulations.
In fact, the right angle scattering phenomenon
seems to be a robust feature in various
two-dimensional models which have soliton solutions 
\cite{zakr,theodora}.
However, it is a nontrivial
and strange behaviour at least from the point of view of
scattering of ordinary particles.

The colliding objects in the two models that we study are essentially
different from each other.
While vortices within
model (\ref{eq:sigma}) are topologically nontrivial objects,
the colliding vortex-antivortex pairs in (\ref{eq:lle}) have
a vanishing topological charge.
However, we argue that 
the underlying Hamiltonian structure allows to study the soliton interaction
in the two models in close analogy.

The outline of the paper is as follows. In Section II 
we simulate head-on collisions of vortex-antivortex pairs 
in (\ref{eq:lle}) and give
a theoretical description which exploits the
form of the linear momentum.
In Section III the results of head-on collision simulations of vortices
in model (\ref{eq:sigma})
are given together with arguments for the
understanding of this behaviour.
The conclusions are given in Section IV.

\section{Head-on collisions of vortex-antivortex pairs in planar ferromagnets}

  A ferromagnet can be described in terms of a magnetization vector
which satisfies the Landau-Lifshitz equation (\ref{eq:lle}).
The constraint on the field $\nbm$ can be resolved and the theory
can be formulated in terms of a complex variable
\begin{equation}\label{eq:omegadef}
 \Omega = \Omega(x,y,t) = {\nm_1 + i\, \nm_2 \over 1 + \nm_3}
\end{equation}
which satisfies the equation
\begin{equation}\label{eq:omegaeq}
 i\, \dot{\Omega} = - \Delta \Omega + {2\, \bOmega \over
1 + \oob}\; \partial_\mu \Omega \; \partial_\mu \Omega
 - {1-\oob \over 1 + \oob}\; \Omega.
\end{equation}
$\bOmega$ denotes the  complex conjugate of $\Omega$.

We use the formulation through the complex variable
$\Omega$ in all numerical simulations. We avoid the
formulation through the vector variable $\nbm$
since the constraint on it makes
an accurate computer calculation of the time derivatives of the field
rather cumbersome.

The model has some interesting static vortex solutions of the form
\begin{equation}
\label{eq:staticvortex}
\Omega^o = f(\rho)\, e^{i\kappa\phi}, \qquad
 \kappa = \pm 1,
\end{equation}
where $\rho, \phi$ are polar coordinates and
$f(\rho=0)=0, f(\rho \rightarrow \infty) \rightarrow 1$.
We call the configuration with $\kappa\!=\!1$ a vortex
and the one with $\kappa\!=\!-1$ an antivortex.
Vortex solutions have infinite energy
and it has been argued that they are physically relevant \cite{gross,afm}.

In the study of the dynamics of magnetic vortices
the central role is played by a scalar quantity called the local vorticity 
\cite{pt,afm}
\begin{equation}
\label{eq:vorticitydef}
 \gamma = \emn\, \partial_\mu \pi \, \partial_\nu \psi,
\end{equation}
where $\emn$ is the two-dimensional totally antisymmetric tensor.
The two components of the linear momentum are then expressed as
\begin{equation}\label{eq:momentum}
 p_x = - \int{y\, \gamma\; dx dy}, \qquad
   p_y = \int{x\, \gamma\; dx dy}.
\end{equation}

Of fundamental importance is the Poisson bracket relation between the two
components of the linear momentum. This reads
\begin{equation}\label{eq:poissonmomentum}
 \{ p_x, p_y \} = \Gamma,
\end{equation}

\begin{figure}
   \begin{center}
   \psfig{file=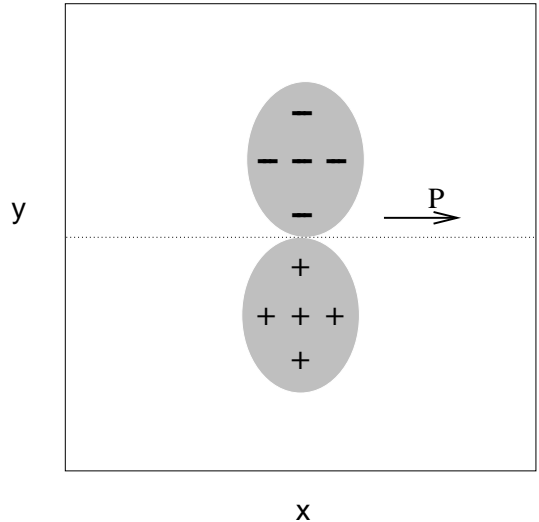,width=7.0cm}
   \end{center}
   \vspace{-5pt}
   \caption{
      A simple vorticity distribution of a
      soliton consists of two lumps with opposite sign.
      They are here symmetrically placed
      on either side of the $x$-axis.
      The lower shaded area represents positive vorticity
      while the upper shaded area represents negative vorticity.
      P denotes the linear momentum of the pair.
   }
   \label{fig:prototype}
\end{figure}
\vspace{10pt}

\noindent where
\begin{equation}\label{eq:totalvorticity}
 \Gamma = \int{\gamma \,dx dy}
\end{equation}
is the total vorticity.

In the present model $\pi\!=\!\cos\!\Theta$ and $\psi\!=\!\Phi$
have been used \cite{pt} as the canonical fields. They are defined through
$n_1\!=\!\cos\!\Theta \sin\!\Phi,\, n_2\!=\!\cos\!\Theta \sin\!\Phi, \,
n_3\!=\!\sin\!\Theta$.
The explicit form of the vorticity is
\begin{equation}
\label{eq:vorticitylle}
 \gamma = \emn\, \sin\!\Theta\;\partial_\nu\Theta \, \partial_\mu\Phi.
\end{equation}
The total vorticity of a vortex is
\begin{equation}
\label{eq:totalvorticitylle}
\Gamma = \int{\gamma \,dx dy} = -2\,\pi\kappa,
\end{equation}
that is, $\Gamma\!=\!-2\,\pi$ for vortices
and $\Gamma\!=\!2\,\pi$ for antivortices.

The implications of a nonvanishing
total vorticity to the dynamics is an issue which has been
thoroughly studied in the case of magnetic vortices and bubbles
\cite{pt,afm,thiele} as well as in the 
case of vortices in other interesting
models \cite{gle,manton}.
The most striking result is that it
leads to spontaneous pinning of these topological objects.

Since a nonvanishing total vorticity  $\Gamma$ implies
pinning of an object, we infer that a solution which
moves freely should have a vanishing $\Gamma$. 
In this respect, the vortex-antivortex ansatz offers the simplest
possibility.
Fig.~\ref{fig:prototype} gives
a schematic representation of it.
This consists of two lumps,
one having negative and the other one positive sign.
We suppose that the vortex is roughly laying
in the shaded area with the negative sign
and the antivortex in the shaded area with the positive sign.
This figure is supposed to act only as a guide for our
discussion and there is no strict way to distinguish the two
vortices and define their positions.
However, if relation (\ref{eq:poissonmomentum}) is applied to each
vortex separately then the quantity on the right hand side
is nonvanishing. It is then implied that
each vortex will propagate in the horizontal direction
under the influence of the other vortex.
The picture is consistent with linear momentum considerations.
That is, an application of Eq.~(\ref{eq:momentum})
to the full ansatz gives a nonvanishing $x$-component of the linear momentum.
Fig.~\ref{fig:prototype} will serve in the following 
discussion as a prototype and
will motivate our theoretical arguments.

The Kelvin motion of a vortex-antivortex pair in a ferromagnet
has been investigated in \cite{mertens}.
The motion of a bubble-antibubble ansatz has also been studied
\cite{papzakr}.
The situation is found to be similar in some other systems such as
an antiferromagnet immersed in a uniform magnetic field \cite{afm},
a model for superconductors \cite{stratos},
for superfluid helium  \cite{roberts}
and the nonlinear Schr\"odinger equation \cite{staliunas}.
It has been pointed out that the gross features of 
this dynamical behaviour are analogous to the planar motion of charges under
the influence of a magnetic field perpendicular to the plane.
In particular, two oppositely charged particles undergo
Kelvin motion traveling along parallel trajectories.
The analogy has been made precise by use of relation
(\ref{eq:poissonmomentum}) and an
analogous relation in the charge motion problem \cite{pt}.

\begin{figure}
   \begin{center}
   \psfig{file=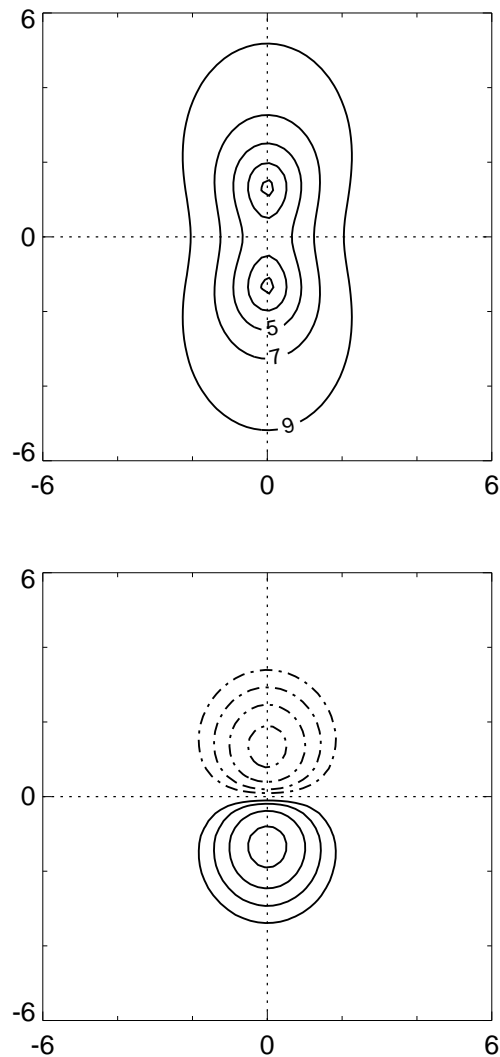,width=6.5cm,bbllx=170bp,bblly=236bp,bburx=390bp,bbury=690bp}
   \end{center}
   \caption{
    Contour plot for a vortex-antivortex pair soliton in a ferromagnet with
    velocity $v\!=\! 0.5$. The upper entry gives contour plots of the
   quantity $10\,|\Omega|$. We plot the levels 1, 3, 5, 7, 9.
   The lower entry is a
   contour plot of the local vorticity for the same soliton.
   Solid lines represent positive values and dashed-dotted lines
   negative values of vorticity.
   We plot the levels $\pm 0.1, \pm 0.2, \pm 0.4, \pm 0.8, \pm 1.2$.
   }
   \label{fig:fmsoliton}
\end{figure}

The calculation of steadily moving coherent structures 
in a 2D ferromagnet, which have
the form of a vortex-antivortex pair has been done in \cite{semi}.
We have used here the numerical code of \cite{semi}
to reproduce them since there is no available analytical formula.
Fig.~\ref{fig:fmsoliton} is an example contour plot for a soliton with
velocity $v\!=\! 0.5$. The upper entry gives contour plots of the
quantity 10 $|\Omega|$ and the two-vortex character of
the configuration is rather obvious. The lower entry is a
contour plot for the local vorticity (\ref{eq:vorticitylle}).
An important result of the analysis in \cite{semi} is that the velocity
is collinear with the linear momentum.

We are now sufficiently motivated to explore the
possibility of scattering of vortex-antivortex pair solitons.
We denote by $\Lambda_v(x,y)$  the solution with
velocity $v$ along the x-axis (set $t\!=\!0$).
 The product ansatz
\begin{equation}\label{eq:ansatzlle}
 \Omega (x,y) = \Lambda_v \left(x\!+\!{\delta \over 2},y\right) \; 
                  \Lambda_{-v} \left(x\!-\!{\delta \over 2},y \right)
\end{equation}
represents two vortex-antivortex pair 
solitons at a distance $\delta$ apart which are in
a head-on collision course.
The ansatz (\ref{eq:ansatzlle}) is used as an initial condition in a
straightforward numerical integration of Eq.~(\ref{eq:omegaeq}). 
We typically set $v\!=\!0.5$, $\delta \!=\! 10$. 

\begin{figure}
   \begin{center}
      \psfig{file=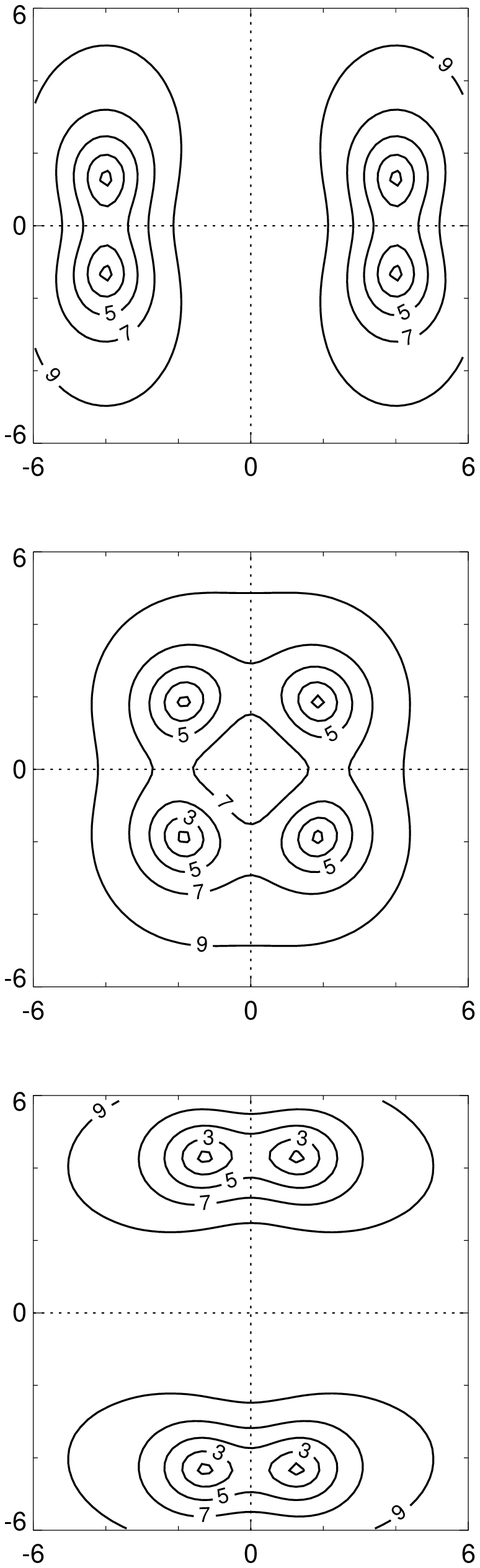,width=6.0cm,bbllx=170bp,bblly=65bp,bburx=390bp,bbury=775bp}
   \vspace{30pt}
   \caption{
   Contour plot of the field $10\,|\Omega|$
   at three characteristic snapshots
   of the head-on collision simulation of vortex-antivortex pair solitons in
   ferromagnets.
   It is shown: the initial ansatz (upper entry, time $t\!=\!0$),
   a snapshot at the time of collision (middle entry, $t\!=\!6.6$),
   and well after collision (lower entry, $t\!=\!13.2$).
   Contour levels as in Fig.~\ref{fig:fmsoliton}.
   }
   \label{fig:fmomega}
   \end{center}
\end{figure}

\begin{figure}
   \begin{center}
    \psfig{file=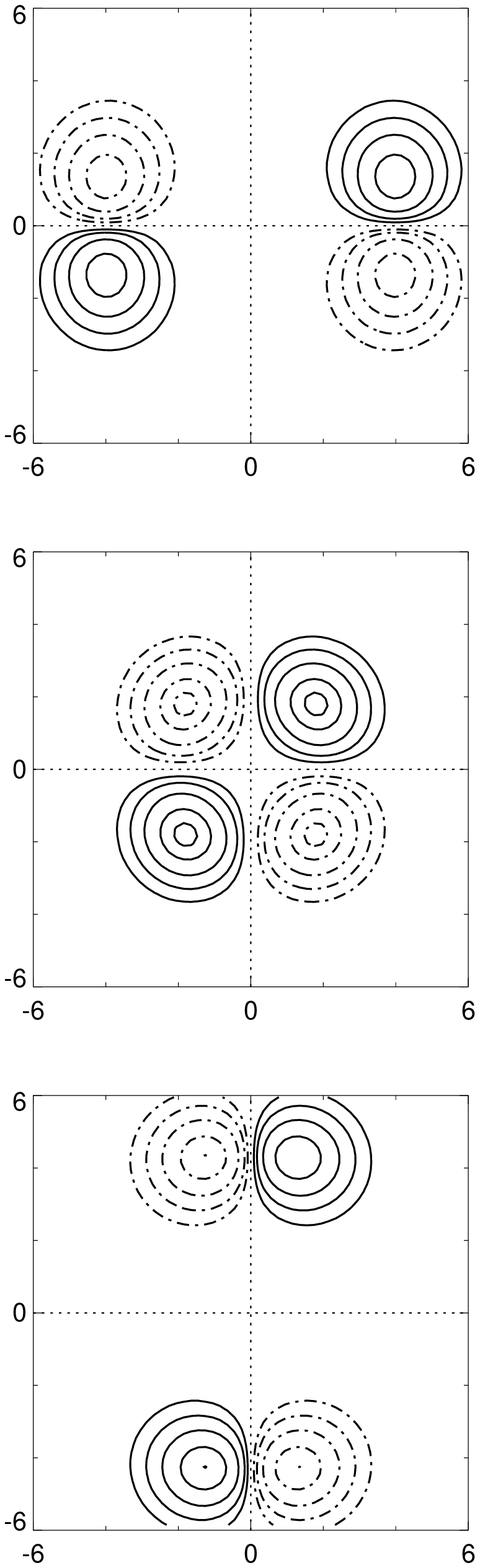,width=6.0cm,bbllx=170bp,bblly=65bp,bburx=390bp,bbury=775bp}
   \vspace{30pt}
   \caption{ Contour plot of the local
   vorticity $\gamma$ of Eq.~(\ref{eq:vorticitylle}) for the solitons of
   Fig.~\ref{fig:fmomega}.
   Contour levels as in Fig.~\ref{fig:fmsoliton}.
   }
   \label{fig:fmvorticity}
   \end{center}
\end{figure}

We have set up a numerical mesh  as large as 600$\times$600 with
uniform lattice spacing $h\!=\!0.1$.
The space derivatives are calculated by finite differences
and the time integration is performed by a fourth order Runge-Kutta
method.
The results are presented
in Figs.~\ref{fig:fmomega} and \ref{fig:fmvorticity}. 
Fig.~\ref{fig:fmomega} presents a contour plot for the field
10 $|\Omega|$ at three characteristic snapshots.
In the first entry the initial ansatz (\ref{eq:ansatzlle}) is shown.
The second snapshot is taken when the solitons
are more or less at a minimum separation.
No vortex-antivortex annihilation process takes place.
This behaviour should be expected since the vortex-antivortex pair
solitons are stable solutions of the equation.
The argument is supported by
numerical simulations showing that a vortex-antivortex
ansatz preserves its character when traveling, provided
that the vortex and antivortex are not very close to each other 
\cite{mertens,afm}.
The last snapshot shows the system after the collision.
A right angle scattering pattern has been produced.

\vspace{10pt}

\vfill \eject

\begin{figure}
\begin{center}
  \psfig{file=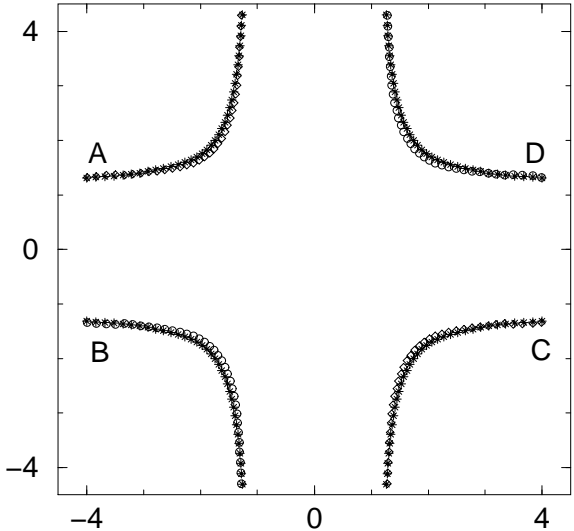,width=7.0cm,angle=-90,bbllx=160bp,bblly=230bp,bburx=420bp,bbury=505bp}
\vspace{25pt}
\caption{Stars denote the zeros of $\Omega$ during the
   numerical simulation of Fig.~\ref{fig:fmomega}.
   We also trace the maximums (circles) and minimums (diamonds)
   of the vorticity. The solitons are initially
   located at $AB$ and $CD$ respectively.
   Symbols are plotted every 0.4 time units.
   }
   \label{fig:fmorbit}
\end{center}
\end{figure}

The situation becomes clearer in Fig~\ref{fig:fmvorticity} where we represent
the solitons in terms of their local vorticity distribution.
The soliton on the left half plane should be compared directly
with that in the lower entry of Fig.~\ref{fig:fmsoliton}. It obviously has
a linear momentum and velocity pointing to the 
positive $x$-direction.
The other soliton has the opposite linear momentum and velocity.

It is rather clear from the picture that the solitons
will not bounce back after collision. This is precluded by the
form of the local vorticity distribution. This possibility 
would require that the vortex and antivortex interchange their position.
Instead, the possibility appears
that, at collision time, the two pairs of vorticity lumps in the
upper and lower half-planes will form two new vortex-antivortex pairs.

One has to apply Eq.~(\ref{eq:poissonmomentum}) 
for each of the vorticity lumps separately.
Alternatively, one can consider pairs of lumps which
tend to travel parallel to each other undergoing Kelvin motion.
Application of this idea to Fig.~\ref{fig:fmvorticity},
determine the time evolution of the system.
Finally, the two pairs on the upper and lower half planes
tend to travel parallel to each other
along the $y$-axis and form bound states.

An equivalent point of view is to follow the linear momentum
of each soliton separately. The linear momentums of the
outgoing solitons clearly lay on the $y$-axis and
have opposite sign.

A subtle but important question is whether we can apply
Eq.~(\ref{eq:poissonmomentum}) separately for each of the
vortices which consist the vortex-antivortex pair.
A rigorous answer can not be given here. On the other
hand the construction of solitary waves in \cite{semi,cooper,roberts}
suggests an affirmative answer whose range of validity
is interesting to study.

The two solitons emerging after collision are very similar
to the initial ones though not exactly the same.
In fact the drift velocity of the outgoing solitons is
somewhat larger.
In Fig.~\ref{fig:fmorbit} we have traced, during the time evolution,
the points where the complex field $\Omega$ vanishes and
also the points where the vorticity $\gamma$ attains its maximum and 
minimum values.
The two kinds of extrema
are close during the whole period of time evolution. This is
because the solitons used in the simulations of this chapter have
a pronounced vortex-antivortex character.
In \cite{piette} traveling solutions of the Landau-Lifshitz equation
have been studied which are different than the ones used
here. Numerical simulations of scattering
of these solitons also produce a right angle pattern.

It is possible to give a picture, corresponding to the
scattering of vortex-antivortex pairs, in terms of 2D motion of
charged particles interacting via their electric field
and placed in a magnetic field perpendicular to the plane.
In fact, we have to consider two electron-positron pairs.
Consider the first electron-positron pair located at points $A, B$
of Fig.~\ref{fig:fmorbit}. and the second pair at $C, D$.
The charges move similar to the vortices.
Their actual orbits resemble
those for the solitons shown in Fig.~\ref{fig:fmorbit}.

Our last remark in this section goes to some related work
in hydrodynamics. There are solutions of the two-dimensional Euler equations
which describe vorticity dipoles. Note that, in this context,
vorticity has its ordinary hydrodynamic meaning.
The best-known such solution
seems to be the Lamb dipole \cite{lamb}. Another one has been found in
\cite{rasmussen}.
A head-on collision between two dipoles 
produces a pattern analogous to that in Fig.~\ref{fig:fmvorticity}
of the present paper \cite{rasmussen,orlandi}.
Furthermore, a simple construction is
given in \cite{lamb} page 223, to which our
Fig.~\ref{fig:fmorbit} can be compared.
Further interesting cases of scattering between 
pairs of objects in hydrodynamics
have been studied. The most complex behaviour has been
observed in \cite{shchur} and includes stochastic and quasiperiodic motion
of vortices.

\section{Head-on collisions of vortices in antiferromagnets}

Our main objective is to study scattering of vortices within
model (\ref{eq:sigma})
and to show that the process can be studied in close analogy 
to the corresponding phenomenon in the ferromagnet.
A right angle scattering behaviour of solitons has been observed
in the isotropic $\sigma$-model \cite{zakr}, 
that is model (\ref{eq:sigma}) without the anisotropy term.

The examination of the local vorticity $\gamma$ has led to a successful 
approach for the collision of vortex-antivortex pairs in ferromagnets
in Section II.
We find it instructive to look at the collision process in terms
of the vorticity also in the present model.
A simple generalization of definition (\ref{eq:vorticitydef}) can be used
\cite{afm}.
The vorticity attains
a simple form when it is expressed in terms of the vector field $\nbm$:
\begin{equation}\label{eq:vorticitysigma}
 \gamma = \emn\; \partial_\mu \dot{\nbm} \cdot \partial_\nu \nbm
        = \emn\; \partial_\mu (\dot{\nbm} \cdot \partial_\nu \nbm).
\end{equation}

In \cite{afm} the equation for an antiferromagnet
in a uniform magnetic field was studied. Vortices in this
system are spontaneously pinned, thus their dynamics is
analogous to that of
ferromagnetic vortices. This unexpected behaviour is probed
by a topological term which enters the vorticity.
However, such a term is absent in the model studied in this section.

The vorticity (\ref{eq:vorticitysigma}) has the form of a total divergence and
can be integrated in all space to show that the total
vorticity vanishes for solutions with reasonable behaviour at infinity:
\begin{equation}\label{eq:totalvorticitysigma}
\Gamma = \int{\gamma\; dx dy} = 0.
\end{equation}
In particular, it vanishes for the vortex solutions.
Relations (\ref{eq:momentum}) - (\ref{eq:totalvorticity}) 
apply in the present context
without modification and they will be the fundamental relations
to be used in the following analysis. 

\begin{figure}
   \begin{center}
   \psfig{file=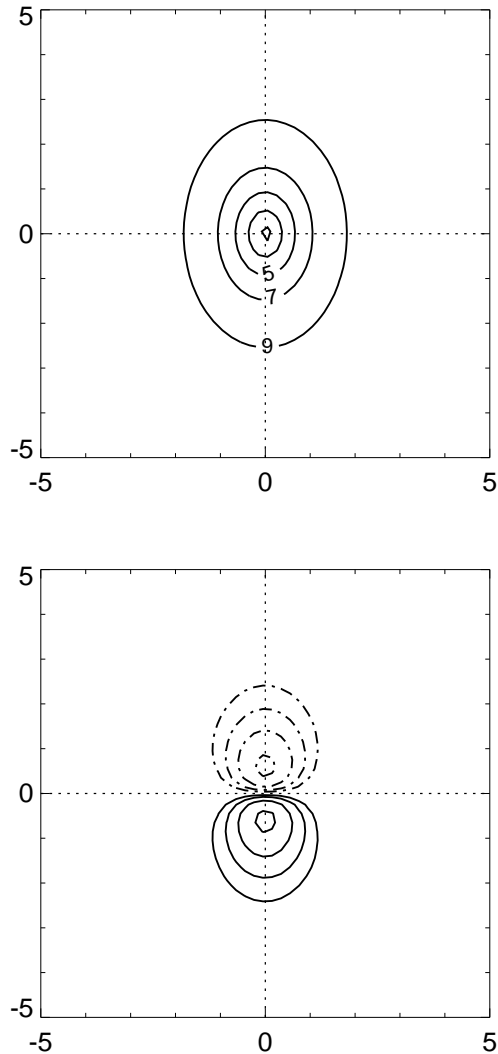,width=6.5cm,bbllx=170bp,bblly=236bp,bburx=390bp,bbury=690bp}
   \end{center}
   \vspace{5pt}
   \caption{Contour plot for a traveling vortex with
   velocity $v\!=\!0.7$. The upper entry gives contours of the
   field $10\,|\Omega|$. The lower entry is a contour plot of
   the local vorticity $\gamma$.
   Solid lines represent positive values and dashed-dotted lines
   represent negative values of vorticity.
   Contour levels as in Fig.~\ref{fig:fmsoliton}.
   }
   \label{fig:vortex}
\end{figure}

Eq.~(\ref{eq:vorticitysigma}) shows that
for a static vortex $\gamma$ vanishes identically.
On the other hand, we can obtain a steadily traveling vortex
by applying a Lorentz transformation to the static
vortex solution (\ref{eq:staticvortex}).
We denote the traveling vortex by $\Omega_v^o$ and the
velocity is $0< v <1$.
The distribution of $\gamma$ for a Lorentz boosted vortex
is nonvanishing and can be calculated numerically.
The vortex with velocity $v\!=\!0.7$ is represented 
by a  contour plot of the field $10 |\Omega|$ 
in the upper entry of Fig.~\ref{fig:vortex}.
A corresponding plot for $\gamma$ is given in the lower
entry of the figure.

The vorticity distribution has the form of two lumps,
thus it resembles the sketch of Fig.~\ref{fig:prototype}. 
This is no surprise. In fact the following two remarks make 
it plausible. Firstly, we see that the total vorticity
vanish according to relation (\ref{eq:totalvorticitysigma}).
Secondly, an inspection of the form (\ref{eq:momentum}) of the
linear momentum makes it clear that a nonvanishing component
is furnished by two lumps of vorticity with opposite signs.
This is not the only form of local vorticity that furnishes a
nonvanishing linear momentum but it is certainly the simplest.
Since the vortex solution with $\kappa\!=\!1$
is indeed the one with the simplest topological complexity,
we expect its vorticity distribution to
have the simplest possible form.

We calculate numerically the points where the maximum and minimum of
the vorticity lumps are located. It turns out that 
these points are the $(0,\pm 0.59)$
for any value of the velocity $v$.

A further example on the present ideas is offered by the
Belavin-Polyakov solutions \cite{bp}.
We apply a Lorentz transformation, with velocity $v$,
to the simplest one:
$ \Omega = {(x - v t)/  \sqrt{1-v^2}} + i y.  $
Its local vorticity is
\begin{equation}\label{eq:bpvorticity}
 \gamma = - {16\; v \over 1- v^2}\; \;
   {y \over \left( 1 + {(x-v t)^2 \over 1-v^2} +y^2 \right)^3}.
\end{equation}
In accordance with the above remarks, it has the shape of two lumps
with opposite sign, located on either side of the $x$-axis
and traveling along the $x$-axis.

\vspace{10pt}

We are now ready to present numerical simulations of head-on collisions of
vortices.
We make an ansatz representing two vortices.
The choice is not unique and the
simplest one seems to be the product ansatz:
\begin{equation}
\label{eq:ansatzsigma}
\Omega(x,y) = \Omega^o \left( x\!+\!{\delta \over 2},y \right)\;\;
                \Omega^o \left( x\!-\!{\delta \over 2},y \right),
\end{equation}
where $\Omega^o$ is the single vortex solution
given in Eq.~(\ref{eq:staticvortex}). 
The two vortices are a distance $\delta$ apart.

The numerical mesh as well as the details of the algorithm that we 
use here are similar to those of Section II.
We use vortices with $\kappa\!=\!1$.
They are initially at rest but immediately
start to drift away from each other due to their mutual repulsion
and escape to infinity.

In order to invoke a head-on collision between vortices we
consider the product ansatz of two vortices which have
opposite velocities:
\begin{equation}\label{eq:ansatzsigma2}
 \Omega(x,y) = \Omega_v^o \left(x+{\delta \over 2},y \right)\;\;
                 \Omega_{-v}^o \left(x-{\delta \over 2},y \right).
\end{equation}
$\Omega_v^o(x,y)$ denotes the Lorentz transformed vortex solution
with velocity $v$, at $t\!=\!0$.

We typically use $\delta\!=\!6$.
In all simulations the vortices start to move 
against each other
with velocities close to the value $v$ but they immediately
begin to decelerate due to their mutual repulsion.
The future of the process depends crucially on
the magnitude of the velocity. At low velocities the two vortices
approach to a minimum distance at which they come to rest and then
turn round and move off in opposite directions. When the velocity
exceeds a critical value (which is $v_c \!\approx\! 0.65$ for $\delta\!=\!6$)
the two vortices collide and scatter at right angles.
This result does not depend on the details of the initial ansatz or on the
initial velocity of the vortices, as long as this exceeds
the critical value. We have tested our algorithm for velocities
up to the value $v\!=\!0.9$ in ansatz (\ref{eq:ansatzsigma2}).
However, one must keep in mind that 
the velocity of the vortices at the time of collision
is smaller than the velocity in the initial ansatz.

\begin{figure}
      \begin{center}
   \psfig{file=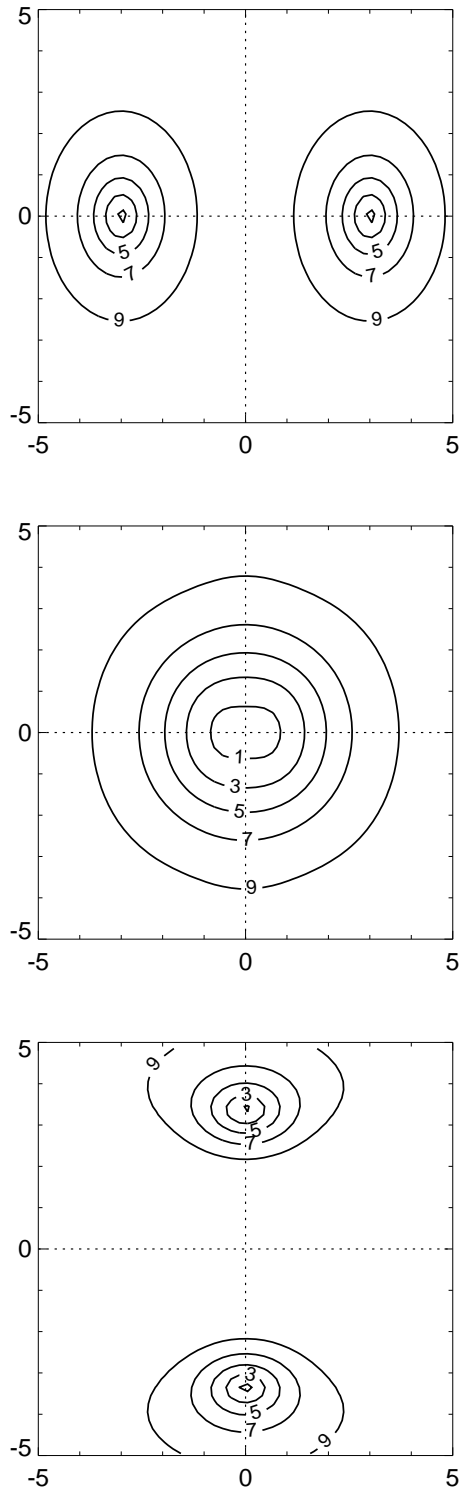,width=6.0cm,bbllx=170bp,bblly=65bp,bburx=390bp,bbury=775bp}
   \end{center}
   \caption{Contour plot of the field $10\,|\Omega|$
   at three characteristic snapshots
   for the head-on collision simulation of vortices in an antiferromagnet.
   It is shown: the initial ansatz (upper entry, time $t\!=\!0$),
   a snapshot at the time of collision (middle entry, $t\!=\!4.2$),
   and well after collision (lower entry, $t\!=\!8.4$).
   Contour levels as in Fig.~\ref{fig:fmsoliton}.
   }
   \label{fig:sigmaomega}
\end{figure}

In Fig.~\ref{fig:sigmaomega}  
we give a contour plot for the norm of the field $\Omega$
at three characteristic snapshots.
The first entry presents the initial configuration (\ref{eq:ansatzsigma2}). 
In the middle snapshot, taken at collision time,
it is clear that the two vortices come
on top of each other. There is no  topological
reason, related to the field $\Omega$, that could prevent this
double vortex to form and there is also no such reason
that could prevent the vortices either to continue
traveling in the horizontal direction or to reemerge traveling in
the vertical direction. We add that, at the present level
of description, we can find no reason that would
enforce them to follow either of the two possibilities.
In the last snapshot the two new vortices that emerge after the collision,
are drifting away from each other along the $y$-axis.

We find it instructive to look at the collision process
using the vorticity. Our description will closely follow that
in Section II in connection with the scattering
of vortex-antivortex pairs.
The dynamics in both systems is determined by the corresponding 
vorticity distribution.
A comparison of the lower entries of
Figs.~\ref{fig:fmsoliton} and \ref{fig:vortex} 
gives a hint that the underlying
dynamics should be of a similar nature in both models.

\begin{figure}
   \begin{center}
   \psfig{file=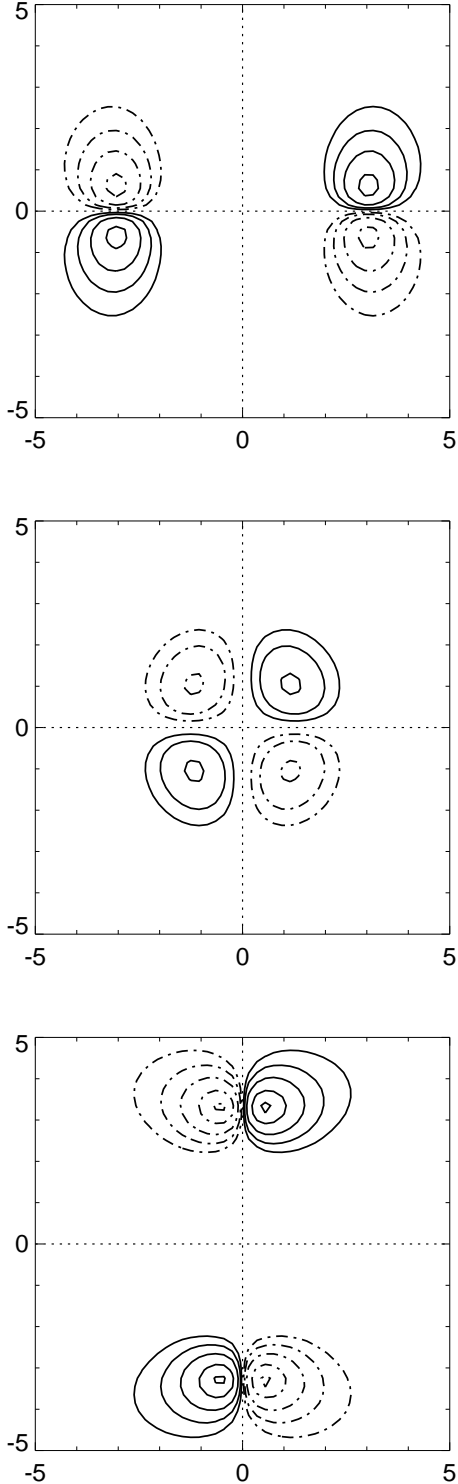,width=6.0cm,bbllx=170bp,bblly=65bp,bburx=390bp,bbury=775bp}
   \end{center}
   \caption{Contour plot of the local
   vorticity $\gamma$ of Eq.~(\ref{eq:vorticitysigma})
   for the vortices of Fig.~\ref{fig:sigmaomega}.
   Contour levels as in Fig.~\ref{fig:fmsoliton}.
   }
   \label{fig:sigmavorticity}
\end{figure}

In Fig.~\ref{fig:sigmavorticity} we give the vorticity
at three snapshots which
correspond to those of Fig.~\ref{fig:sigmaomega}.
Fig.~\ref{fig:sigmavorticity} should be compared directly 
with Fig.~\ref{fig:fmvorticity}.
An examination of these results shows that the arguments
of Section II for the soliton scattering which rely upon the
linear momentum relations (\ref{eq:momentum}), 
(\ref{eq:poissonmomentum}) are applicable here, too.

The upper entry in the figure corresponds to the initial ansatz
and each of the two vorticity dipoles should be compared
to that given in the lower entry
of Fig.~\ref{fig:vortex}.
The two vortices in the first entry of Fig.~\ref{fig:sigmavorticity} 
are approaching each other while their
dynamical features, as described by the vorticity, are not substantially
modified. The repulsion which could decelerate them
and make them turn round, is overcompensated by the large
enough initial velocity.
When the two vortices come close to each other
(middle entry) the two pairs of 
vorticity lumps, lying in the upper and lower half plane, interact.
The subsequent evolution of the vorticity lumps is governed
by relation (\ref{eq:poissonmomentum}). 
In particular, this relation has to be applied
to each of the vorticity lumps separately. As is indicated by the
simulation, the lumps survive throughout the process and the
simple dynamics implied by (\ref{eq:poissonmomentum}) 
is sustained during the scattering process.
Following the discussion in Section II we examine the dynamics of lumps 
in a pairwise manner.
The two pairs in the upper and lower plane tend to move
along the $y$-axis. Consequently, the initial partners separate and
two new vortices are formed 
which travel in opposite directions along the $y$-axis,
as shown in the lower entry of the figure.

An equivalent approach is to follow the linear momentum of the
solitons.
The two pairs in the lower and upper half-plane have their linear momentums
lying along the $y$-axis but with opposite signs.
They subsequently tend to go off along the $y$-axis.
The important numerical result 
is that the vorticity lumps roughly preserve their shape 
throughout the process. This is due to the fact that traveling vortices
are stable solutions of the model.

We note here that our result is obtained when 
Eq.~(\ref{eq:poissonmomentum}) is applied to each vorticity lump
separately. We have been motivated to follow this approach
because of its success with respect to studying the dynamics
of vortex-antivortex pairs in a ferromagnet and because of the
consistency of the picture with the linear
momentum considerations. On the other hand
a rigorous proof of its validity is lacking.

\begin{figure}
   \begin{center}
   \psfig{file=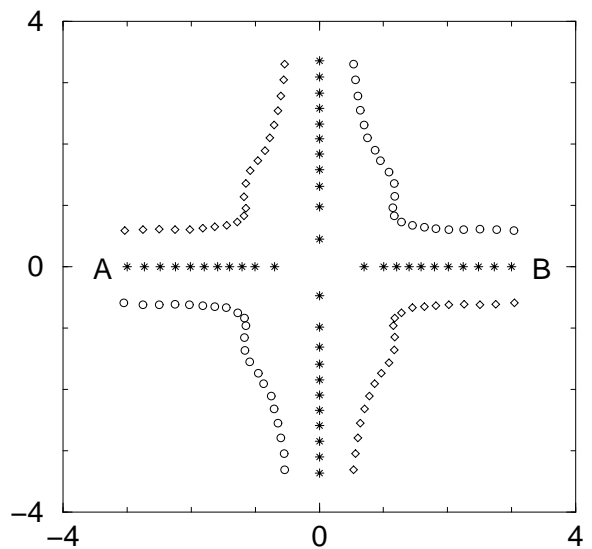,width=7.0cm,angle=-90,bbllx=160bp,bblly=230bp,bburx=420bp,bbury=505bp}
   \end{center}
   \caption{Stars denote the zeros of $\Omega$ during the
   numerical simulation of Fig.~\ref{fig:sigmaomega}.
   We also trace the maximums (circles) and minimums
   (diamonds) of the vorticity. The vortices are initially
   located at points $A$ and $B$.
   Symbols are plotted every 0.4 time units.
   }
   \label{fig:sigmaorbit}
\end{figure}
\vspace{10pt}

In Fig.~\ref{fig:sigmaorbit} we track some characteristic 
points of the vortices
throughout the simulation time. The stars denote successive
points where the centers
of the two vortices lie during the simulation,
that is, where the complex function $\Omega$ vanishes.
The circles denote successive locations of the 
maximum and the diamonds the locations of the minimum of the
vorticity distribution. We plot
two circles and two diamonds at every time instant. 
At the beginning of the simulation the two vortices are
centered at points $A$ and $B$ respectively. They immediately start to
decelerate but when their centers reach a distance
$\approx  2.4$ they seem to accelerate considerably, they 
eventually merge at the
origin and then they separate along the $y$-axis.
The trajectories of the extrema of the vorticity show that after
the collision it takes some time until the two new vortices
are organized again.
The velocity at the end of the
numerical simulation is somewhat lower than that in the initial ansatz.
This should be due to spin waves emitted during the process.

The remarks of the last paragraph on the motion
of the vortex centers are in agreement with an analytical solution obtained
in \cite{wardomega} representing scattering of solitons in an
integrable chiral model.
Close to collision time, the centers
move according to the law $x \approx \pm \sqrt{-t}$
$(t < 0)$.
This gives a velocity $dx/dt \rightarrow \infty$
as $t \rightarrow 0$. They collide at $t\!=\!0$
and the centers of the two new solitons,
which emerge along the $y$ axis, follow $y \approx \pm \sqrt{t}\; (t>0)$.

The arguments presented in this section are certainly not sufficient
to exclude, other than right angle, interesting
possibilities of interaction and scattering of solitons.
Nevertheless, they imply that the right angle scattering
process is expected to be generic for solitons in two-dimensional
Hamiltonian models.
In the present work we have found no relation of the topology
of the solitons to the right-angle scattering phenomenon,
Therefore right-angle scattering is also expected to occur among
non-topological solitons \cite{leandros}.

\section{Conclusions}

We have given a description of the right angle phenomenon of solitons
through numerical simulations, as well as arguments which suggest
that it should be generic in two dimensions.
Two systems have been examined. Vortex-antivortex pairs in
planar ferromagnets and vortices in antiferromagnets.
The peculiar scattering behaviour is mainly attributed
to the fact that solitons are extended structures rather than
point like particles. This point is accounted for
by the representation of a traveling soliton
through a pair of lumps.
Furthermore, we find that these two lumps act as independent physical entities
at the time of collision.

It is desirable to observe experimentally scattering
of solitons in ferromagnetic and antiferromagnetic films.
We expect that the present theoretical analysis
will be useful in studies of systems of a lot of vortices
\cite{mertens} and in particular in studies of the thermodynamics of magnetic
systems. Suffice it to say that, in a magnetic material,
vortices are expected to appear in pairs.
In the study of the thermodynamics of layered antiferromagnets
one expects to find the signature of topological excitations.
The remark may prove important especially in view of the
difficulty to observe isolated antiferromagnetic solitons
due to the lack of a significant total net magnetization.

The right angle scattering pattern appears also and has been
understood from the point of view of the geometry of the moduli space
for BPS monopoles in a three-dimensional
Yang-Mills-Higgs theory \cite{hitchin}.

We have proceeded in the
simulation of scattering of a vortex and an antivortex
in the $\sigma$-model (\ref{eq:sigma}).
We use an initial ansatz similar to that of Eq.~(\ref{eq:ansatzsigma}).
We simulate the time 
evolution of the system numerically and find
that the vortices attract each other. They eventually collide
at the origin and annihilate. It is quite interesting that the energy
is dissipated at right angles. The phenomenon is presumably closely related to
the present study.
A similar simulation with corresponding results has been
done in \cite{shellard}  for non-gauged cosmic strings.

The basic dynamical structure
which leads to right angle scattering in Hamiltonian models
is also present in the complex Ginzburg-Landau equation (CGLE) 
which describes nonlinear oscillatory media \cite{kramer}.
There is actually a variety of nonconservative systems where
scattering behaviour analogous to that studied in the present paper 
has been observed.
An interesting example is a 2D fluid layer subjected to
externally imposed oscillations. Coherent structures are formed
which are experimentally observed to scatter at an almost right angle
\cite{lioubashevski}.

\section{Acknowledgments}
I am grateful to N. Papanicolaou and P.N. Spathis
for providing me their numerical code which calculates the
vortex-antivortex pairs used in the simulations of Section II
and for useful discussions. 
I thank F.G. Mertens for discussion of the issues presented in this paper.
I also thank L. Kramer for beneficial remarks on part of the text
and L. Perivolaropoulos for his help with the
numerical algorithm.


\widetext

\end{document}